\documentclass{emulateapj}
\usepackage{lscape} 
\newcommand{\ha}{H$\alpha$}
\newcommand{\msun}{$M_{\odot}$}
\newcommand{\vi}{$V-I$}
\newcommand{\mv}{$M_{V}$}

\begin{document} 

\title{The Chromospheric Activity and Ages of M Dwarf Stars in Wide
Binary Systems\altaffilmark{1}}

\shorttitle{Activity and Ages of M Dwarfs} 
\shortauthors{Silvestri et~al.}
\slugcomment{Accepted for publication in the Astronomical Journal, May 2005}

\author{Nicole M. Silvestri\altaffilmark{2}, Suzanne
L. Hawley\altaffilmark{2}, and Terry D. Oswalt\altaffilmark{3}}

\altaffiltext{1}{Based on observations obtained with the Apache Point
  Observatory (APO) 3.5-meter telescope, which is owned and operated
  by the Astrophysical Research Consortium (ARC); The Cerro Tololo
  Inter-American Observatory (CTIO) 4.0-meter telescope, which is
  operated by the Association of Universities for Research in
  Astronomy (AURA) Inc., under a cooperative agreement with the
  National Science Foundation (NSF) as part of the National Optical
  Astronomy Observatories (NOAO), which also operates Kitt Peak
  National Observatory in Tucson, Arizona; and the SARA Observatory
  0.9-meter telescope at Kitt Peak, which is owned and operated by the
  Southeastern Association for Research in Astronomy ({\tt
  http://www.saraobservatory.org}).}

\altaffiltext{2}{Department of Astronomy, University of Washington,
Box 351580, Seattle, WA 98195, nms@astro.washington.edu,
slh@astro.washington.edu.}

\altaffiltext{3}{Department of Physics \& Space Sciences and the SARA
Observatory, Florida Institute of Technology, 150 W. Univ. Blvd.,
Melbourne, FL 32901, toswalt@fit.edu.}

\begin{abstract}

We investigate the relationship between age and chromospheric activity
for 139 M dwarf stars in wide binary systems with white dwarf
companions.  The age of each system is determined from the cooling age
of its white dwarf component.  The current limit for activity-age
relations found for M dwarfs in open clusters is 4 Gyr. Our unique
approach to finding ages for M stars allows for the exploration of
this relationship at ages older than 4 Gyr.  The general trend of
stars remaining active for a longer time at later spectral type is
confirmed.  However, our larger sample and greater age range reveals
additional complexity in assigning age based on activity alone.  We
find that M dwarfs in wide binaries older than 4 Gyr depart from the
log-linear relation for clusters and are found to have activity at
magnitudes, colors and masses which are brighter, bluer and more
massive than predicted by the cluster relation.  In addition to our
activity-age results, we present the measured radial velocities and
complete space motions for 161 white dwarf stars in wide binaries.

\end{abstract}


\section{Introduction} 
\setcounter{footnote}{4}

The study of stellar activity has long been restricted to observed
features on the surface of the Sun and to bright solar-type stars
\citep{Bya,BD}.  Studies conducted by, among others, \citet{Wil66},
\citet{Sku}, and \citet{Noyes} showed a link between activity in
solar-type stars and the stellar rotation rate.  An $\alpha \Omega$
dynamo driven by differential rotation at the radiative-convective
boundary layer is thought to be the primary driver behind stellar
activity in F, G, K, and early M stars.  It is also believed to be
responsible for the heating of the stellar chromosphere and corona as
well as the production of star spots, flares, and prominences.

Recently, the study of stellar activity has been extended to include
the entire M dwarf sequence.  At first glance the magnetic activity
observed in these low mass stars appears similar to that found in
solar type stars, and in some cases is particularly strong.  The
dynamo generation mechanism must be different however, as these stars
should be fully convective and therefore lacking the
radiative-convective boundary layer responsible for magnetic dynamo
generation in more massive stars \citep[cf.][]{PH,ONS,DF,MA}.

Due to their slow burning of core hydrogen, M dwarfs that formed at
the birth of the Galaxy are still on the main sequence and therefore
retain information from the Galaxy's earliest epochs of star
formation.  In this study we investigate the relationship between
magnetic activity and age for M dwarfs, with the goal of developing an
age-activity relation that can be used to determine ages for this
important and ubiquitous population.

Stellar age is one of the most difficult properties of a star to
determine.  \citet{Sku} showed that chromospheric activity and age are
related for solar-type stars by:
%%%%%%%%%%%%%%%%%%%%%%%%%%%% 
\begin{equation} 
F({\rm Ca\ K}) = At^{-1/2}, 
\end{equation} 
%%%%%%%%%%%%%%%%%%%%%%%%%%%% 
where $F(\rm Ca~K)$ is the equivalent width of the Ca~{\sc II}~K
chromospheric emission line (in \AA), $A$ is a constant, and $t$ is
the age of the star (in years).  The common explanation for this
effect is that as the star ages, its rotation slows due to mass and
angular momentum loss. This leads to a weakening of the internal
rotationally-driven dynamo, causing a decrease in magnetic heating and
hence chromospheric activity. Calibration of the chromospheric
activity$-$age relation for F, G, and K stars has been carried out by
\citet{Barry} and \citet{SDJ}.  However, a similar relation does not
appear to hold for the M dwarfs.  \citet{S91} originally showed that
all stars redder (lower temperature, lower mass) than a given $R-I$
color had H$\alpha$ emission in the Pleiades and Hyades clusters.  The
color at the ``H$\alpha$ limit" was bluer, corresponding to a higher
mass where activity still occurred, in the younger cluster.  This
result led to the speculation that both mass and age affect the
H$\alpha$ limit.  A subsequent study of six open clusters
\citep[IC2602, IC2391, NGC2516, Pleiades, Hyades, and M67][hereafter
HRT]{HRT} showed that there was a linear relationship between the \vi\
color at the \ha\ limit and the log(age) of the cluster.  The \vi\
color is directly related to the atmospheric temperature and therefore
the mass of a main sequence M dwarf. The relationship was calibrated
only for relatively young M dwarfs with ages $<$ 4 Gyr (the age of
M67, the oldest open cluster in their study).  Thus, HRT confirmed
that activity in M dwarfs is a function of age at a given mass (or
alternatively mass at a given age), quite unlike the
rotationally-dependent \citeauthor{Sku} relation for F$-$K stars,
which appears to be independent of mass.  The HRT age-activity
relationship was used by \citet{GRH} to constrain the ages of brown
dwarf stars in binaries with M dwarf companions and to study the star
formation history of the Galactic disk.

Unfortunately, it is not yet feasible to spectroscopically determine
the activity levels of older M dwarfs in either open or globular
clusters, because the clusters are faint and distant, making the
observations difficult even on the largest telescopes.  Therefore, the
current activity$-$age relation determined with the use of clusters
only allows for the sampling of stars formed well after the earliest
epochs of star formation in the Galaxy. Further progress in pinning
down the old, low mass end of the M dwarf chromospheric activity$-$age
relation must therefore come from another method.

Here we describe our analysis of a sample of $\sim$ ~200
spectroscopically identified M dwarf stars in common proper motion
binary (CPMB) systems with white dwarf (WD) companions.  The age of
the system is determined from the cooling age of the companion WD,
assuming the WD and M dwarf are coeval.  The WD ages are accurate to
roughly 25\%\ (or better if the mass of the WD is known), comparable
to the accuracy determined from the ages of clusters.  Since many
binaries in our sample have ages well beyond the present 4 Gyr cluster
limit, this allows us to examine the activity$-$age relation for much
older M dwarfs. Concurrently, our sample provides a significant amount
of new data on WD stars in binary systems.

Using the spectra of the M dwarfs, we determine the spectral types
(masses) and from the photometry of the WDs we determine the age of
each pair in which activity is observed.  In addition, we determine
the radial velocities of the M stars and, since proper motions are
known, the complete space motions for all systems in the observed
sample.  This allows us to examine the kinematics and provides an
independent indicator of population membership.

The paper is organized as follows: In \S2 we introduce the sample and
selection criteria, and describe the observations, data reduction and
determination of M dwarf atmospheric parameters. Our method for
estimating ages of the WDs using atmospheric and evolutionary models
is presented in \S3.  In \S4 we examine the spectroscopic and
photometric properties of the active M dwarfs and present the final
activity$-$age results for the sample.  We discuss the kinematics of
the sample and the implications of these results in \S5, building on
the earlier work in \citet{SOH}. Our conclusions are summarized in
\S6.

\section{Sample Selection and Observations}

\citet{Luy63,Luy69,Luy74,Luy79} and \citet{GBT71,GBT78} first
identified a sample of over 500 wide binary systems in several proper
motion-selected surveys from the original Palomar Observatory Sky
Survey (POSS) plates and the photographic plates from Lowell
Observatory.  The Luyten$-$Giclas CPMB systems have relatively wide
orbital separations ($\langle a \rangle \sim 10^{3}$ A.U.,
\citealp{O93}).  Unlike close binary systems, these wide pairs are not
influenced by mass exchange and evolve essentially as two single
stars.  The stars that make up a CPMB (or multiple component) system
originally evolved from the same molecular cloud
\citep{Boss87,Boss88,Pri} and are therefore taken to be coeval. Of
particular interest to our study are those CPMBs which contain one WD
and one main sequence M dwarf star (we will refer to these systems as
WD+dM pairs).  Among these pairs are some with very faint and cool
(hence old) WD stars \citep{LDM,OSWH}, which are particularly useful
for setting a firm lower limit to the age of the Galaxy.  Here we also
show that they are useful for extending the age range of the
age-activity relationship.

A subset of 196 systems was selected for observation (see
Table~\ref{all}).  The sole selection criterion was that the system
must contain spectroscopically identified WD and M dwarf stars.  Low
resolution spectra (7$-$15 \AA\ pixel$^{-1}$, \citealp{OHL,OSHL,O93})
were used to confirm the rough spectral type of each CPMB
component. Of the 196 systems, 189 are binary (WD+dM) and seven are
triple systems (one WD+dM+dK, one WD+dM+dG, three WD+dM+dM, and two
WD+WD+dM).  This gives a total of 199 M dwarf stars with at least one
WD companion.  Table~\ref{all} lists all 196 system names as assigned
by Luyten in column 1, their R.A. and Decl. (coordinates are for
equinox 1950; columns 2 \& 3), the $V$ magnitude and original low
resolution spectroscopic identification (columns 4-7), the site, date,
and exposure time at which the object was observed (columns 8-10).
Columns 11-14 list the proper motion, direction of proper motion
(measured east of north), position angle (centered on the primary
measured east of north), and the separation of the components,
respectively.

The biases in our sample, imposed primarily by the sampling area,
proper motion ($\mu$) criteria, and limiting apparent magnitude (in
this case photographic magnitude, $m_{\rm pg}$) of the original
surveys, are well understood \citep[see][]{OS,SO,OSWH,WO}.  Our sample
includes objects north of declination $-60\degr$ excluding the most
dense regions of the Galactic plane ($\mid$b$\mid$ $<$ 15$\degr$).  It
thus includes $\sim$ 65\% of the visible sky \citep{Daw}.  The
Luyten$-$Giclas binaries are restricted to 0.1 $\leq \mu \leq$
2$\arcsec$.5 yr$^{-1}$ \citep{La}, a limit based solely on Luyten's
ability to discern motion between the original and retake plates of
the POSS.  The magnitudes of the CPMBs are $m_{\rm pg}$ $\leq$ 20,
i.e.  down to the rough magnitude limit of the POSS blue plate.  It
should also be noted that because Luyten avoided listing pairs that he
knew had already been discovered by Giclas, the latter need to be
included in any sample where completeness is important.

\subsection{Spectroscopic Observations}

The northern portion of the sample was observed spectroscopically
during 32 half-nights on the ARC 3.5-m telescope at APO.  Observations
were performed on the Double Imaging Spectrograph in high-resolution
mode ($\sim$~2~\AA/pixel) with a 1$\arcsec$.5~slit.  With this setup,
slits were centered at 6700~\AA\ (red) and 4600~\AA\ (blue) yielding
wavelength coverage of $\sim$6000-7500~\AA\ and $\sim$4000-5200~\AA,
respectively.  An additional three full nights of observing time on
the CTIO 4-m Blanco telescope were used to observe the systems which
could not be seen from the northern hemisphere. Observations were
performed with the R$-$C Spectrograph in medium-resolution mode
($\sim$ 4~\AA/pixel) and a 1$\arcsec$.0 slit. The grating tilt was
centered at 6800~\AA\ and covered a wavelength range of
$\sim$3500-8500~\AA.  A condensed Journal of Observations is given in
Table~\ref{lo} where the first three columns give the UT date of the
observing run followed by the site (column 4), the average seeing for
the night (column 5), and some notes on the weather conditions (column
6).

We were unable to observe ten (40Eri$-$C, LP36$-$142, LP369$-$14,
LP321$-$397, LP321$-$459, LP458$-$48, CD-51$\degr$13128, L427$-$61,
LP411$-$22, and L573$-$109) of the 196 systems in our original sample
(see Table~\ref{all}).  40Eri$-$C was too close to its extremely
bright companion to be observed.  LP36$-$142, LP369$-$14, LP321$-$397,
and LP321$-$459 were all too faint ($m_{\rm pg}$ $\geq$ 19.5) to be
observed in reasonable exposure times.  The proper motion of the
system LP458$-$49/48 has moved the M star (LP458$-$48) into the glare
of a bright field star.  CD$-$51$\degr$13128, L427$-$61, LP411$-$22,
and L573$-$109 were not observed due to weather on the evenings on
which they were slated for observation.

The data were reduced with standard $IRAF$\footnote{$IRAF$ is written
and supported by the $IRAF$ programming group at the NOAO in Tucson,
Arizona.  NOAO is operated by the AURA, under cooperative agreement
with the NSF ({\tt http://iraf.noao.edu/}).} reduction procedures.  In
all cases, program objects were reduced with calibration data (bias,
flat, arc, flux standard) taken on the same night. Data were bias
subtracted and flat-fielded, and one-dimensional spectra were
extracted using the standard aperture extraction method.  A wavelength
scale was determined for each target spectrum (including the flux and
M dwarf standards) using He-Ne-Ar arc lamp calibrations. Flux standard
stars were used to place the spectra on a calibrated flux scale. We
emphasize that the final flux calibrations for the targets are only
relative fluxes as most nights were not spectrophotometric.

After final reductions and close inspection of the individual spectra,
we eliminated 28 M dwarfs from the final analysis because of
insufficient signal to noise in the spectra. An additional 22 binaries
were eliminated due to problems with the WD photometry as outlined in
\citet{Smith}.  This left 139 WD+dM binaries for final analysis in the
activity$-$age relation in \S4 and 161 M dwarf stars for the
kinematics discussion in \S5 (kinematics of the system are based on
the M star so the 22 systems with poor WD photometry are included in
this portion of the study).

\subsection{Spectroscopic Measurements}

\begin{figure}[hbt]
\plotone{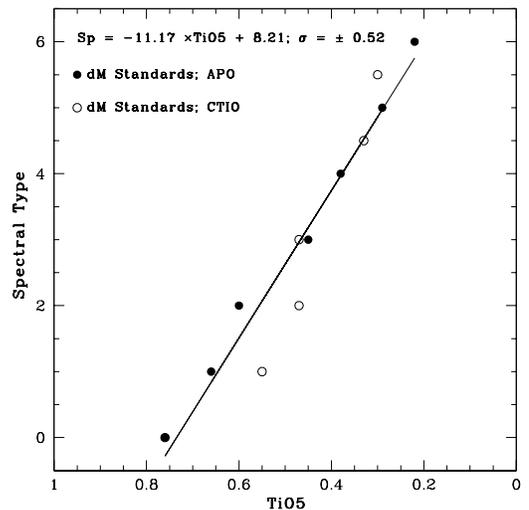}
\caption{The measured TiO5 bandhead strength versus catalog spectral
type \citep{RHG}.  The straight line denoted by the equation on the
plot is the adopted relation for our sample over the range dM0 $\leq$
Sp $\leq$ dM6. The filled circles are standards observed at APO with
$\sim$ 2 \AA/pixel resolution and uncertainty of $\pm$ 0.5 spectral
type.  The open circles are from the lower resolution spectra observed
at CTIO with $\sim$ 4 \AA/pixel resolution and indicate an uncertainty
of $\pm$ 1.0 spectral type. \label{tiorel}}
\end{figure}

The spectral types of the M dwarfs were determined by measuring the
strength of Titanium Oxide (TiO) molecular features in the spectra,
using flux ratios at strong bandheads as described in \citet{RHG}.  We
used observations of M dwarf standards obtained at APO to calibrate
the spectral type scale for our program as described in \citet{KHM}
and \citet{H20}, and found that our results agreed well with
\citet{RHG}.  The uncertainty in M dwarf spectral type for this
procedure is $\pm$ 0.5 spectral type at the APO resolution and $\pm$
1.0 spectral type at the CTIO resolution. Figure~\ref{tiorel} plots
our measured TiO5 bandstrengths against the catalog spectral type.  As
displayed in the Figure the equation for the line of best fit is
%%%%%%%%%%%%%%%%%%%%%%%%%%%%%%%%%%%%%%%%%%%%%%%%%%%%%%%%%%%%%%%%%%%%%%%%%%%
\begin{equation} 
\rm Sp = -11.17 \rm \times TiO5 + 8.21, \sigma_{\rm rms} = \pm 0.52.
\end{equation}
%%%%%%%%%%%%%%%%%%%%%%%%%%%%%%%%%%%%%%%%%%%%%%%%%%%%%%%%%%%%%%%%%%%%%%%%%%%
The relation adopted for this (TiO5, Sp) calibration by \citet{RHG}
for their sample of 88 M dwarf standards defined by \citet{KHM} is
%%%%%%%%%%%%%%%%%%%%%%%%%%%%%%%%%%%%%%%%%%%%%%%%%%%%%%%%%%%%%%%%%%%%%%%%%%%
\begin{equation}
\rm Sp = -10.775 \rm \times TiO5 + 8.2, \sigma_{\rm rms} = \pm 0.5.
\end{equation}
%%%%%%%%%%%%%%%%%%%%%%%%%%%%%%%%%%%%%%%%%%%%%%%%%%%%%%%%%%%%%%%%%%%%%%%%%%%
Clearly, the relation is in good agreement with the larger sample of
standards from \citet{RHG}.

% FIGURE 1 %%%%%%%%%%%%%%%%%%%%%%%%%%%%%%%%%%%%%%%%%%%%%%%%%%%%%%%%%%%%%%%
%%%%%%%%%%%%%%%%%%%%%%%%%%%%%%%%%%%%%%%%%%%%%%%%%%%%%%%%%%%%%%%%%%%%%%%%%%%

Calcium Hydride (CaH) features were also measured, as the relative
strength of CaH compared to TiO serves to identify subdwarfs as
described in \citet{RG}.  We found only one possible subdwarf in the
sample, LP164-52.

The equivalent width (EW) of the H$\alpha$ emission line, which is
formed in the magnetically heated chromosphere and is commonly used to
characterize the magnetic activity of M dwarfs, was measured following
the prescription in \citet{HGR}.  The H$\alpha$ EW may change
dramatically through the M dwarf sequence solely because of the
rapidly changing continuum flux in this wavelength region, and not
because of a change in the magnetic heating of the chromosphere.
However, all of our analysis is carried out on subsamples of objects
with the same spectral type, so comparing relative equivalent widths
within these subsamples gives a reasonable measure of differences in
magnetic activity.

\begin{figure}[hbt]
\plotone{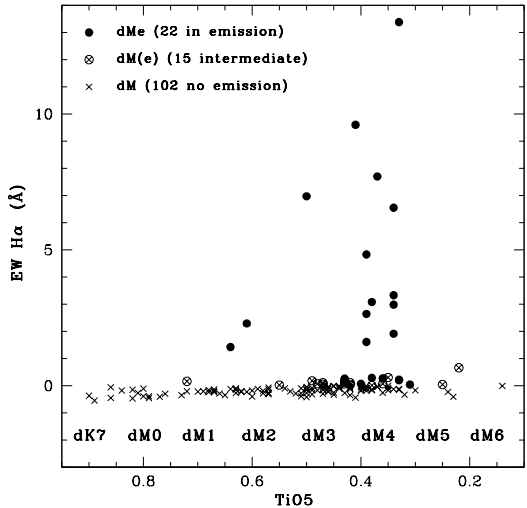}
\caption{The EW in \AA\ of the \ha\ feature for 139 dMs in CPMBs
versus the ratio of the flux in the TiO5 molecular band.  Filled
circles are dMe stars. Open circles with crosses are dM(e) stars and
crosses are inactive dM stars. \label{EWratio}}
\end{figure}

Table~\ref{ew} gives the measured H$\alpha$ EW and uncertainties
(columns 4-6) for each of the 139 M dwarfs. Calcium hydride and TiO
flux ratios are listed in columns 7-14 followed by the new spectral
types as determined from the TiO5 bandheads (column 15). A reference
column (16) indicates the designations (dMe, dM(e), or dM) for each M
dwarf as discussed further in \S 4.1.  The H$\alpha$ EW and TiO5
ratios are displayed in Figure~\ref{EWratio}.

% FIGURE 2 %%%%%%%%%%%%%%%%%%%%%%%%%%%%%%%%%%%%%%%%%%%%%%%%%%%%%%%%%%%%%%%
%%%%%%%%%%%%%%%%%%%%%%%%%%%%%%%%%%%%%%%%%%%%%%%%%%%%%%%%%%%%%%%%%%%%%%%%%%%

\section{Determination of White Dwarf Ages}

We use the photometry of the WDs provided by \citet{Smith} for our age
determinations.  This method introduces somewhat larger uncertainties
in age than matching model H line profiles to high S/N data
\citep[e.g.][]{Sil2} but it is accurate enough for our purposes
\citep[see][for a discussion of this procedure]{OSWH,Smith}.  The
cooling ages were extracted from the model grids as described in
\citet{BSW} using the $V-I$ and $B-V$ colors of the WDs.

The mass of the WD plays the dominant role in determining the WD
cooling rate.  Hence, it is not surprising that a large uncertainty in
the final age from the WD models is associated with the mass estimate.
Unfortunately, determining the mass of a WD is not a simple task, and
achieving a high precision is even more daunting.  At present, there
are less than 300 individual WDs for which a mass has been determined
and of those, less than 100 have been determined within $\sim$~10\%
precision ($\sigma <$ 0.05 \msun).

\citet{Sil2} determined the masses for 44 WDs in the Luyten sample
based on the gravitational redshift of the WD.  These masses on
average have $\sigma_{M_{\rm WD}}$ $>$ 0.05 \msun. \citet{Sil2}
sampled a significant fraction of the WDs in our current sample.  We
use their mean mass of 0.61 $\pm$ 0.16 \msun\ as a representative mass
of the WDs in our sample.  Only three of the WDs in this study have
independently measured masses and active (dMe) M dwarf companions
(L170-14B, LP617-35, and LP798-13) and when these masses are used in
the determination of the ages for these three systems, the ages are
very close (within the quoted age uncertainty) to the ages determined
from the mean mass.  For consistency, we use the mean mass estimate
for all of the WDs in this study.  It appears to be a reasonable mean
mass assumption since it is close to the mean mass of nearly every WD
mass distribution in the literature \citep[see for
example][]{K87,BSW,Reid}.

Having established an average mass and uncertainty for the WDs in our
sample, we next determined the gravities corresponding to our mean
mass and error using the mass$-$radius relation implicit in the WD
model calculations.  For either the hydrogen- or helium-dominated
model atmospheres, the average difference in cooling ages are less
than 0.5 Gyr, approximately half the size of the errors imposed by
assuming a mean mass for the WD. Adding errors from the photometry and
from the mass estimate in quadrature yields the uncertainty in the
cooling ages of the WDs for the pure H and He (DA and DB WDs,
respectively) atmospheric model results.  The average cooling age for
our sample is $t_{\rm cool}$ = 3.06$^{+0.94}_{-1.09}$ Gyr for a 0.61
\msun\ mass WD.

\subsection{Initial$\rightarrow$Final Mass Relation for White Dwarfs}

The ages and uncertainties calculated above represent the total
cooling age of each individual WD (total time after emerging from the
planetary nebula).  This however, is not the total age of the WD.  We
need a way of estimating the length of time a 0.61~\msun\ WD exists on
the main sequence before evolving to become a WD. The pre-WD lifetime
is dependent on the initial mass of the progenitor star.  Therefore,
it is necessary to determine the main sequence mass of the progenitor
star based on the final WD mass.  This process is complicated because
the progenitor may lose a significant amount of mass during its
evolution.

\citet{W00} discusses the many factors which determine the final mass
of the WD such as rotation, binarity, magnetic fields, mergers, and
differential mass loss. The average main sequence mass for a 0.61
\msun\ WD from \citet{W00} is 2.17 $\pm$ 0.19 \msun. Using the
third-order polynomial of \citet{IL}
%%%%%%%%%%%%%%%%%%%%%%%%%%%%%%%%%%%%%%%%%%%%%%%%%%%%%%%%%%%%%%%%%%%%%%%%
\begin{equation}
{\rm log} t_{\rm evol} = 9.921 - 3.6648({\rm log} M_{\rm I}) +
1.9697({\rm log} M_{\rm I})^{2} - 0.9369({\rm log} M_{\rm I})^{3}
\end{equation}
%%%%%%%%%%%%%%%%%%%%%%%%%%%%%%%%%%%%%%%%%%%%%%%%%%%%%%%%%%%%%%%%%%%%%%%%%%%%
we determine the main sequence lifetime corresponding to a given
initial mass, where $t_{\rm evol}$ is the main sequence lifetime (in
years) of the star and $M_{\rm I}$ is the progenitor mass of the WD
(in \msun).

For our $M_{\rm WD}$ = 0.61 \msun\ WD mass, this translates into a
main sequence lifetime of 0.75$^{+0.21}_{-0.15}$ Gyr.  We added this
value to the individual cooling ages of the WDs ($t_{\rm cool}$) as
determined in the previous section.  This yields an average age for
the WDs in our sample of 3.81$^{+0.98}_{-1.09}$ Gyr, near the age of
the oldest cluster (M67) in the previous activity$-$age relations
determined by HRT.

\section{Age-Activity Results}

\subsection{H$\alpha$ Emission}

In \S2.3 we discussed how the EW of the \ha\ emission feature in M
dwarfs is used as the primary marker of an active star.  For
comparison with recent EW studies, we define a feature with positive
EW as being in emission and negative EW as being in absorption.
Previous studies such as \citet{RHM,HGR,HRT,GRH} use a cutoff of 1.0
\AA\ as the minimum EW at which one can unambiguously detect emission
in a M dwarf. We measured 23 stars with 0.0 $<$ EW $<$ 1.0 \AA\ and 14
stars with EW $>$ 1.0 \AA.  Of the 23 low-emission objects, emission
was confirmed by eye for only eight stars.  In the remaining 15
low-emission stars, the \ha\ feature is too small to distinguish from
the noise.  We have decided to include only the eight stars where \ha\
can be confirmed by eye in our group of M stars with emission (dMe).
Though this is a subjective way of establishing a minimum limit of
detectability, it ensures that we follow only stars that are
definitely in emission through our analysis.  In Figure~\ref{EWratio},
the 22 dMe stars are plotted as filled circles.  The 15 stars with
questionable emission are plotted as open circles with crosses.  We
refer to these as dM(e) or intermediate dM stars, and keep track of
their location throughout our analysis, in an attempt to deduce their
true nature and assess the reliability of detecting emission in M
dwarfs with EW $<$ 1.0 \AA.  Finally the stars that were identified as
having no \ha\ emission (dM) are plotted as crosses.  We maintain this
notation in all subsequent plots.  Although some of the dMs may be
dM(e) within our uncertainty, we are confident that none of the dM or
dM(e) are dMe.

\begin{figure}[hbt]
\plotone{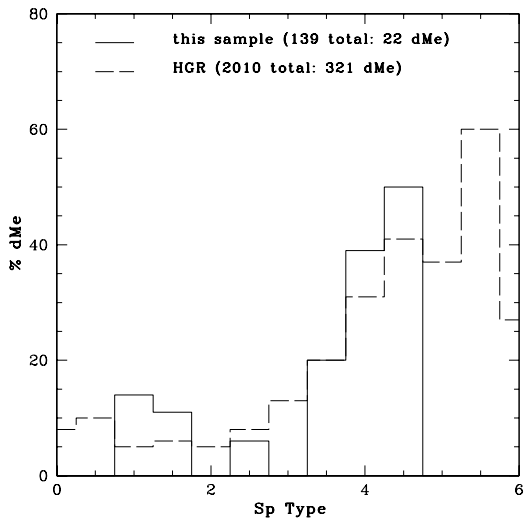}
\caption{Ratio of dMe stars to the total number of M stars in each 0.5
spectral class bin.  The solid line represents our sample, the dashed
line indicates the HGR sample.  There are 139 total stars in our
sample and 2010 total stars in the HGR sample. As discussed in the
text, our sample have very few stars later than type
M4.5. \label{percent}}
\end{figure}

Figure~\ref{percent} illustrates the fraction of stars in emission per
half-spectral class in our sample (solid line) as compared to
\citet[][hereafter HGR - dashed line]{HGR}.  Though the number of
stars per bin is significantly less for our sample than for HGR the
trend toward a higher fraction of dMe to dM stars per 0.5 spectral
type is similar (and within the uncertainty) up to a spectral type of
$\sim$ M4.5 where the number of stars in our sample drops
precipitously.  Only a small fraction of stars are in emission at
early M spectral types in both samples.

% FIGURE 3 %%%%%%%%%%%%%%%%%%%%%%%%%%%%%%%%%%%%%%%%%%%%%%%%%%%%%%%%%%%%%%%%
%%%%%%%%%%%%%%%%%%%%%%%%%%%%%%%%%%%%%%%%%%%%%%%%%%%%%%%%%%%%%%%%%%%%%%%%%%%

\subsection{Activity and Age}

\begin{figure}[hbt]
\plotone{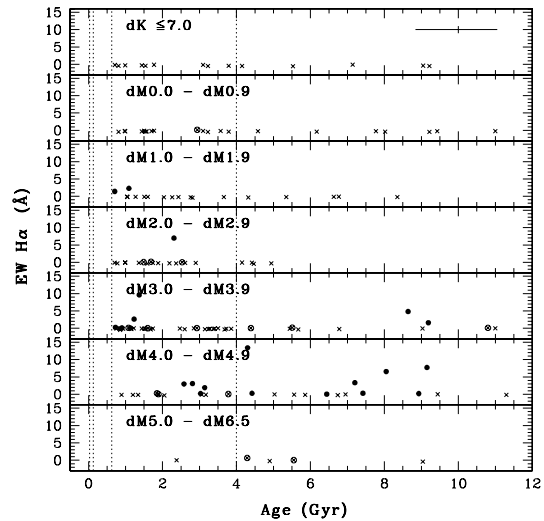}
\caption{The EW \ha\ per 1.0 spectral type bin versus age for our
sample.  The filled circles represent the 22 dMe in our sample.  The
open circles with crosses represent the 15 intermediate dM(e) stars
and finally the crosses, all at or below zero EW, are the remaining
102 inactive dMs.  Vertical dotted lines indicate the ages derived for
clusters observed in the PMSU study.  Average error bars are given at
the right hand side of the top panel.  See text for detailed
discussion. \label{ewage}}
\end{figure}

In Figure~\ref{ewage} we divide the sample by spectral type into seven
separate panels.  The symbols are the same as given in
Figure~\ref{EWratio}.  The top panel contains the earliest spectral
type (late K; previously classified as early M dwarfs based on
low-resolution spectra) stars in the sample.  The next five boxes are
spectral types M0, M1, M2, M3, and M4 respectively.  The last box
contains all of the later type M dwarfs in our sample.  The H$\alpha$
EWs of the stars are plotted (in \AA) versus the ages (in Gyr).

% FIGURE 4 %%%%%%%%%%%%%%%%%%%%%%%%%%%%%%%%%%%%%%%%%%%%%%%%%%%%%%%%%%%%%%%%
%%%%%%%%%%%%%%%%%%%%%%%%%%%%%%%%%%%%%%%%%%%%%%%%%%%%%%%%%%%%%%%%%%%%%%%%%%%%

The average error in age (largely a result of the errors in photometry
and mass) is approximately $\pm$ 1.0 Gyr as shown at the right of the
top panel.  The actual average errors for the EWs are much smaller
(formally $\sigma_{\rm H\alpha}$ = $^{+0.03}_{-0.02}$ \AA) and arise
primarily from the placement of the pseudo-continuum points.  The four
vertical dotted lines represent the ages of the clusters used in the
HRT activity$-$age relation.  The youngest clusters, IC2602 and
IC2391, are both $\sim$ 30 Myr old, while the slightly older clusters
NGC2516 and the Pleiades are grouped together at 125 Myr.  The third
line at 625 Myr is the age of the Hyades cluster and the final line at
4 Gyr is at the age of M67, the oldest cluster in their sample.

There are several important features that warrant discussion in this
Figure.  First, and not surprisingly, it appears that at early ages
($<$ 4 Gyr), there are active M dwarfs in nearly every well-populated
spectral type bin with the exception of M0.  A large fraction of our
dM(e) stars are located within this age range as well.  HRT found that
nearly every star in their youngest clusters was active above a
certain threshold so it is not surprising that a large fraction of our
active stars are apparently young.

However, the majority of our sample is quite old in comparison to the
clusters studied by HRT.  We have no stars as young as their four
youngest clusters and have very few M dwarf stars that are even as
young as the Hyades (625 Myr).  The young end of the cluster
activity$-$age relation is well sampled; as expected, our sample will
contribute important information on activity at relatively old ages. A
large fraction of our stars are older than the oldest cluster (M67 at
4 Gyr) available for the cluster activity$-$age relation.

Another interesting aspect of Figure~\ref{ewage} is the striking lack
of active stars with ages greater than 4 Gyr and spectral types
earlier than M3.  There are virtually no active M stars earlier than
type M3 that are older than roughly 2 Gyr.  Stars in our sample with
spectral types M3 and later appear to remain active for much longer.
Again this agrees qualitatively with the cluster results.

\begin{figure}[hbt]
\plotone{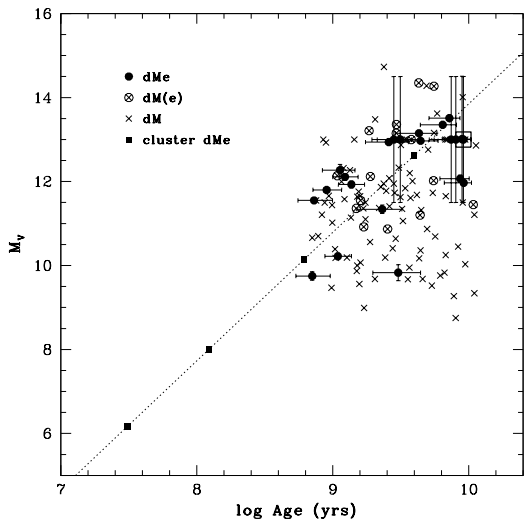}
\caption{The \mv\ versus log(age) relation for our sample.  The points
are the same as in previous figures with the addition of the four
solid squares representing the cluster ages from the PMSU study along
with the corresponding empirical fit (dotted line). The open square
denotes the location of LP133$-$373/374. \label{mvage}}
\end{figure}

\begin{figure}[hbt]
\plotone{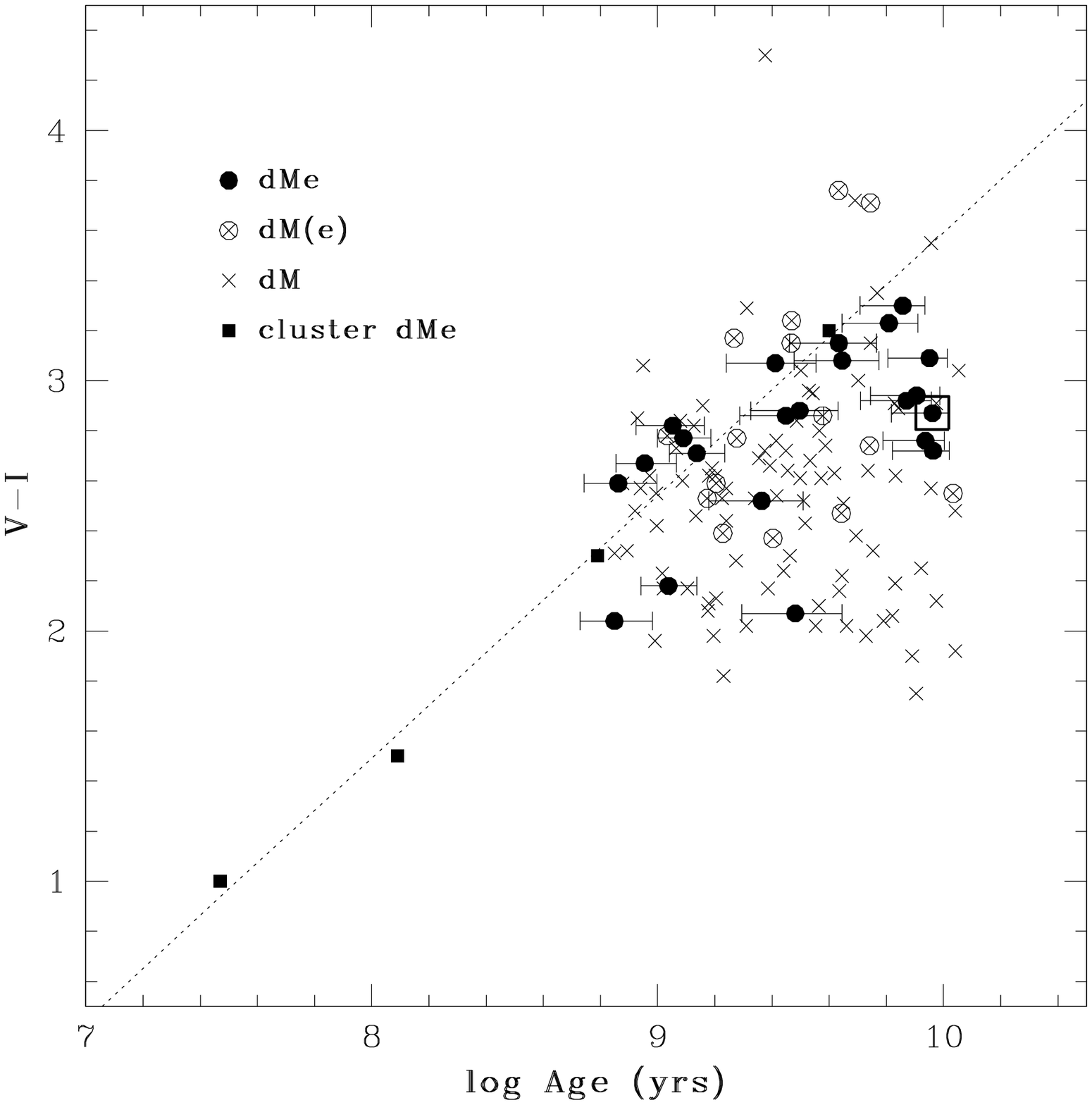}
\caption{The \vi\ versus log(age) relation for our sample. Symbols are
the same as in Figure~\ref{mvage}. The open square denotes the
location of LP133$-$373/374. \label{viage}}
\end{figure}

\begin{figure}[hbt]
\plotone{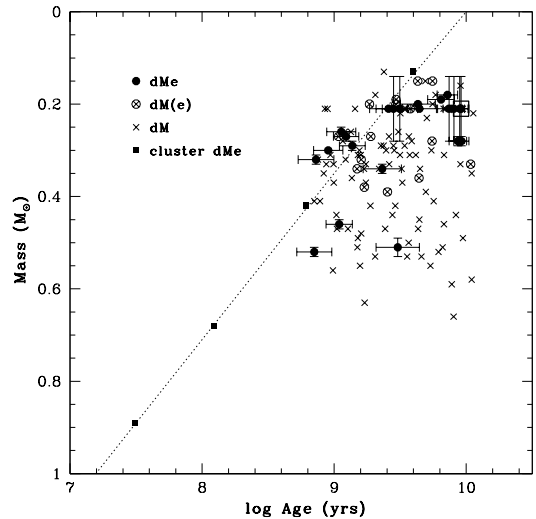}
\caption{The M dwarf mass versus log(age) relation for our sample.
Symbols are the same as in Figures~\ref{mvage} and \ref{viage}. The
open square denotes the location of LP133$-$373/374. \label{massage}}
\end{figure}

The next three figures (Figures~\ref{mvage}-\ref{massage}) show \mv,
\vi, and mass versus log(age) respectively. Table~\ref{vimv} lists the
\vi, \mv, and mass values (columns 4-6) for the active (dMe and dM(e))
stars in Figures~\ref{mvage}-\ref{massage}.  The \vi\ colors for the M
dwarfs in our sample are from \citet{Smith}.  We determined the \mv\
for the stars in our sample based on \citet{RH} 4$^{\rm th}$ order
polynomial fit relating the \mv\ to the \vi\ color of the star as
%%%%%%%%%%%%%%%%%%%%%%%%%%%%%%%%%%%%%%%%%%%%%%%%%%%%%%%%%%%%%%%%%%%%%%%%%%%%%
\begin{equation}
M_{V} = 3.98 + 1.437(V-I) + 1.073(V-I)^{2} - 0.192(V-I)^{3}.
\end{equation}
%%%%%%%%%%%%%%%%%%%%%%%%%%%%%%%%%%%%%%%%%%%%%%%%%%%%%%%%%%%%%%%%%%%%%%%%%%%%%
This polynomial proves to be a good fit, in particular to stars with
0.85 $<$ (\vi) $<$ 2.85.  There is an abrupt discontinuity in the
slope of $M_{V}$ versus \vi\ for dM stars at a \vi\ $\sim$ 2.9 (dM4).
Stars at this point have magnitudes $\pm$ 1.5 in \mv\ in either
direction with the midpoint at \mv\ $\sim$ 13.  For stars with \vi\
$\sim$ 2.9 in our sample, we assigned the midpoint value of \mv\
$\sim$ 13 with errors of $\pm$ 1.5. For all other stars in
Figures~\ref{mvage}-\ref{massage} the uncertainties in \mv, \vi, and
mass are the size of the data points.  For clarity, the uncertainties
in age are given for only the active stars.  The average age
uncertainty for the rest of the points is $\pm$ 1 Gy.

% FIGURE 5 %%%%%%%%%%%%%%%%%%%%%%%%%%%%%%%%%%%%%%%%%%%%%%%%%%%%%%%%%%%%%%%%%
%%%%%%%%%%%%%%%%%%%%%%%%%%%%%%%%%%%%%%%%%%%%%%%%%%%%%%%%%%%%%%%%%%%%%%%%%%%%

% FIGURE 6 %%%%%%%%%%%%%%%%%%%%%%%%%%%%%%%%%%%%%%%%%%%%%%%%%%%%%%%%%%%%%%%%%
%%%%%%%%%%%%%%%%%%%%%%%%%%%%%%%%%%%%%%%%%%%%%%%%%%%%%%%%%%%%%%%%%%%%%%%%%%%%

% FIGURE 7 %%%%%%%%%%%%%%%%%%%%%%%%%%%%%%%%%%%%%%%%%%%%%%%%%%%%%%%%%%%%%%%%%
%%%%%%%%%%%%%%%%%%%%%%%%%%%%%%%%%%%%%%%%%%%%%%%%%%%%%%%%%%%%%%%%%%%%%%%%%%%%

The M dwarf masses were determined using the relation from \citet{Delf}:
%%%%%%%%%%%%%%%%%%%%%%%%%%%%%%%%%%%%%%%%%%%%%%%%%%%%%%%%%%%%%%%%%%%%%%%%%%% 

$\log \frac{M}{M_{\odot}} = 10^{-3}[0.3 + 1.87(M_{V}) + $

$7.6140(M_{V})^{2} \ - 1.6980(M_{V})^{3} \ + \ 0.060958(M_{V})^{4}],  
M_{V} \in [9, 17]. $ \begin{flushright}(6)\end{flushright}
%%%%%%%%%%%%%%%%%%%%%%%%%%%%%%%%%%%%%%%%%%%%%%%%%%%%%%%%%%%%%%%%%%%%%%%%%%% 

Figures~\ref{mvage}-\ref{massage} are directly comparable to the
cluster results shown in HRT, and to Figure~11 (\vi\ plot only) in
\citet{GRH}.  We reproduce the cluster results on our plots with the
dotted line connecting the four filled squares (representing the 4
unique \ha \ turn-off ages of the six clusters). The cluster
age-activity relation identifies the dotted line as the suggested
activity limit: stars that are fainter in absolute magnitude, redder
in color, and lower in mass at a given age than the value of the
relation (i.e. above the line in our plots) should be active, while
stars below the line should be inactive.  Results for our CPMB systems
are shown with the same symbols as in previous plots.

Many of the dMe stars in our sample are more active than the cluster
relation would predict.  The same is true for some of the dM(e) stars.
However, a significant number of the dMe stars fall below the
relation, indicating that with the larger age spread and the sampling
of individual variations possible using our CPMB sample, the cluster
relation is to simplistic.  At the oldest ages we sample, near 10
Gyrs, the dMe stars are often bluer, more massive and, to a lesser
extent, brighter than the cluster relation would predict.  In other
words, magnetic activity is lasting longer in these higher mass stars
than is observed in clusters. The average errors in \mv\ ($<$ 0.5
magnitudes), \vi\ ($<$ 0.1 magnitudes), and mass ($<$ 0.1 \msun)
cannot account for all of the discrepancy.

We note that one possible source of bias in our sample is the presence
of a close, unresolved binary companion.  In early M dwarfs, the
induced synchronous (fast) rotation of such a system can increase the
dynamo heating and produce strong emission even in old systems.
\citet{GRH} point to a few M dwarfs in the D region of their Figure~5
as being active due to the presence of spectroscopically identified
binary companions.  \citet{FM} show that among binaries with M dwarf
components, the incidence of detectable unresolved tertiary components
is about 7\%. \citet{Smith} detected evidence of unresolved tertiary
components (via anomalies in $BVRI$ colors) in about 10\% of the 500
Luyten-Giclas pairs.  Given that binaries comprise about 0.5-0.75 of
the stars in the solar neighborhood \citep{Abt}, one would
expect a sample of single stars to be several times more
seriously impacted by unresolved binary pairs than a wide binary
sample is impacted by unresolved tertiary components.

To investigate this effect, \citet{O04} report on time series
photometric observations of all of our pairs which contain active M
dwarf components in order to search for light variations in either the
WD or main sequence components.  Each pair receives at least a half
night of observing time.  So far, only one, LP133-373/374 (dM4e+DC,
black open square in Figures~\ref{mvage}-\ref{massage}) has shown any
periodic variability that exceeds 2\% in amplitude.  The dMe component
in this pair exhibits partial eclipses with a period of about 19.5
hours, suggesting that its enhanced activity is at least partially due
to tidal interaction between two unresolved nearly identical dMe
stars.  The lightcurves of LP133-373/374 and other wide pairs will be
published elsewhere.  Nevertheless, the low incidence of short term
photometric variability in the CPMB sample suggests that third
components in, general, do not appear to account for most of the
scatter seen in Figures~\ref{mvage}-\ref{massage}.

An interesting finding, which again differs from the cluster results,
is that for our binaries, not all M dwarfs that lie above the limiting
relation are in emission (inactive M dwarfs are marked as crosses in
the Figures).  All (of the observed) cluster M dwarf stars that were
redder (fainter, less massive) than the limiting relation had
H$\alpha$ in emission.  This may indicate that stars in clusters are
more uniform in their activity evolution than those in the field at
the same age. However, again the possibility of unresolved companions,
and/or differences in evolution between M dwarfs in clusters and M
dwarfs in binary systems with a white dwarf may be causing unknown
effects on the magnetic field generation in the M dwarf. As the
driving mechanism behind the magnetic heating of the mid-M dwarfs is
not well understood, there is no way of predicting the effect of
different kinematical or residual orbital effects on such a mechanism.
Evidently variables other than mass and rotation rate may influence
the magnetic activity in such old systems.  Also, we emphasize that
the binaries are from a field star sample with consequent
uncertainties in their individual evolution and properties, which
contribute to the more complicated nature of their activity$-$age
relation.

In summary, the general trend of lower mass (redder, fainter) stars
remaining active longer than was predicted from the cluster results is
confirmed.  However, the scatter is larger, the activity is not
ubiquitous, and the relation is not linear with log(age) for the ages
older than 1 Gyr which we sample with our binaries.

\section{M Dwarf Kinematics}

In this section we examine the kinematics of active and inactive M
dwarfs for the sample of 161 M dwarfs in CPMB systems. The sample
includes an additional 45 M dwarfs observed after the original
investigation of \citet[][hereafter SOH]{SOH}.  Twenty-two of these
stars were excluded from the age-activity analysis in the previous
section because we did not have photometry for their white dwarf
companions and therefore could not determine their ages.  However,
they are included here because we do have good M dwarf data to
determine their space motions. Table~\ref{uvw} gives the new velocity
data for the sample, including the 45 WD+dM systems not presented in
SOH.  Columns 1-3 list the WD name assigned by Luyten, the R.A. and
Decl. (coordinates are for equinox 1950).  Columns 4-7 list the proper
motion, the direction of proper motion, our measured radial velocity
($v_{r}$), and uncertainty in radial velocity ($\sigma_{\rm v_{r}}$).
Columns 8-15 give the $UVW$-space motions, their uncertainties, and
the full space motion of each system. As with previous tables, a
reference column (16) indicates the designation (dMe, dM(e), or dM)
for each M dwarf.  As in SOH, the space velocities are in a
left-handed coordinate system.

\begin{figure}[hbt]
\plotone{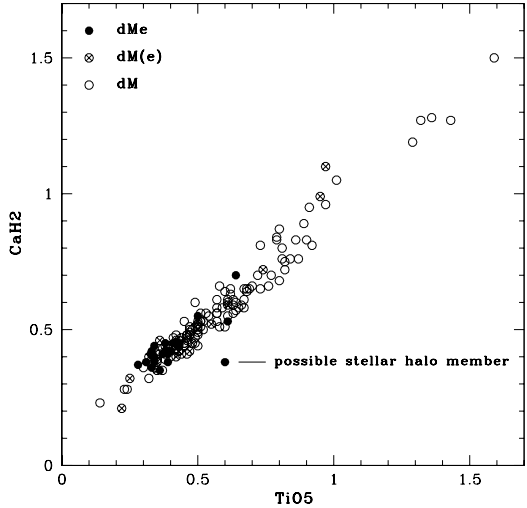}
\caption{The CaH2 versus TiO5 relation for 161 M dwarfs in binary
systems.  The bandpasses for CaH2 and TiO5 are 6814-6846~\AA\ and
7126-7135~\AA, respectively.  Symbols are the same as in
Figures~\ref{mvage}-\ref{massage} except the dMs are now shown as open
circles.  We find only one potential subdwarf system,
LP164-52. \label{halo}}
\end{figure}

The metallicity of the M dwarfs, determined by comparing CaH and TiO
band strengths provides additional information on the population
membership of the CPMB systems.  Figure~\ref{halo} shows the CaH2
versus TiO5 plot originally calibrated by \citet{Gizis} to give a
rough metallicity estimate for M dwarfs. Metal poor halo stars would
lie below and to the right of the disk sequence represented by the
bulk of the points plotted in the Figure.  LP164$-$52 remains the only
probable subdwarf in this sample.  The rest appear to have solar-like
metallicities which suggests that all but one of the 161 binaries are
part of the Galactic disk population\footnote{The two objects in the
upper right hand portion of Figure~3 of SOH have since been reobserved
with higher quality spectra. The new data indicate better agreement
with the rest of the sample in Figure~\ref{halo}.}.

% FIGURE 8 %%%%%%%%%%%%%%%%%%%%%%%%%%%%%%%%%%%%%%%%%%%%%%%%%%%%%%%%%%%%%%%%%
%%%%%%%%%%%%%%%%%%%%%%%%%%%%%%%%%%%%%%%%%%%%%%%%%%%%%%%%%%%%%%%%%%%%%%%%%%%%

\begin{figure}[hbt]
\plotone{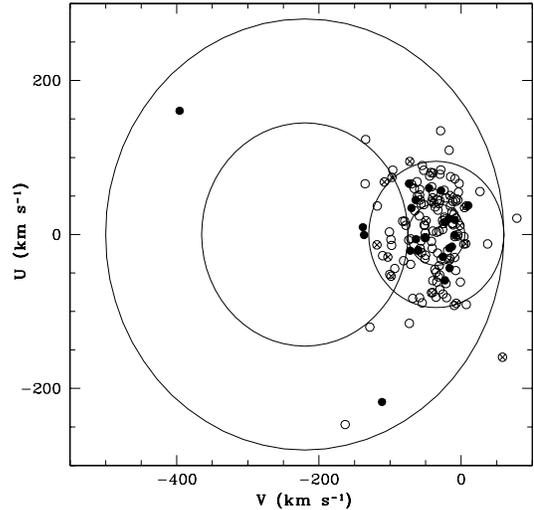}
\caption{The $U$ versus $V$ velocity distribution of 161 binary
systems with measured \ha\ EWs and radial velocities. The symbols are
the same as in Figure~\ref{halo}. The ellipsoids plot the 1$\sigma$
(inner) and 2$\sigma$ (outer) contours for the Galactic thick-disk and
stellar halo populations, respectively. Typical errors are
approximately $\pm$ 10 km~s$^{-1}$. \label{uv}}
\end{figure}

Figure~\ref{uv} shows the $U$- versus $V$-velocity components of the
space motion for the 161 binaries with good M dwarf spectra.  The
symbols are similar to those in previous figures, with dMe stars
denoted by filled circles, dM(e) stars by open circles with
crosses. The dM stars are now represented by open circles for clarity
(they were denoted by crosses in previous Figures). As expected from
the metallicity results, the stars are nearly all contained within the
velocity contours of the Galactic disk (thin and thick components,
including the eclipsing M dwarf, LP133$-$373 at [+19.0, $-$9.0]).
Except for a few obvious outliers (the potential subdwarf, LP164$-$52
at [$-$0.7,$-$136.6], HZ 43B, LP219$-$78, and LP205$-$28), the active
stars have lower average velocities and are more closely confined to
the thin disk, in agreement with previous results \citep{RHG,HGR}. The
marginally active dM(e) stars, in contrast, are more closely aligned
with the higher velocity, inactive dM stars. The plot also reveals a
few additional high velocity systems compared to the earlier SOH
results. The spectra for these high velocity systems are of equivalent
quality to the rest of the spectra and there are no obvious systematic
errors which would result in velocities that are larger than the
average of the distribution.  Their large velocities are likely a
result of being part of the high velocity tail of the thin or thick
disk population of the Galaxy. The additional data shown in
Figure~\ref{uv} support the conclusions of SOH that the high velocity
white dwarf stars in this sample are not part of a dark matter halo,
in contrast to the suggestion of \citet{Op01}.

% FIGURE 9 %%%%%%%%%%%%%%%%%%%%%%%%%%%%%%%%%%%%%%%%%%%%%%%%%%%%%%%%%%%%%%%%%
%%%%%%%%%%%%%%%%%%%%%%%%%%%%%%%%%%%%%%%%%%%%%%%%%%%%%%%%%%%%%%%%%%%%%%%%%%%%%

\section{Conclusions}

We have determined the ages for 139 M dwarf stars in CPMBs to an
accuracy of $\pm$~1~Gyr and have extended the chromospheric
activity$-$age relation to ages much older than the oldest open
cluster used in previous work.  As found in previous work, lower mass
(redder, later spectral type) M dwarfs are more likely to be active at
old ages than higher mass M dwarfs.  However, we have demonstrated
that our activity$-$age results for M dwarfs in wide binary systems
depart from the log-linear relations seen in cluster M dwarfs.
Neither the uncertainties in the age estimates for the binary M dwarfs
nor the uncertainties in cluster age estimates can account for the
disparity between the two relations.  Active M stars of a given age in
our sample are found at higher mass, bluer color and brighter absolute
magnitude than predicted from the cluster results.

In addition, the activity in our CPMB M dwarfs is not pervasive above
the \ha\ limit in color, magnitude, and mass as is found to be the
case for M dwarfs in clusters.  Evidently the magnetic activity
evolution in clusters is more uniform than in these field binaries, at
least for ages $<$ 4 Gyr.  The same activity behavior we observe may
be present in older clusters, which are as yet too faint to be
observed.  It is not clear what mechanism would selectively allow for
emission at a particular age in some old M dwarf stars while
prohibiting it in others of the same mass and spectral type at the
same age.  Possible explanations for the discrepancy include
unresolved close companions that induce fast rotation in some M dwarfs
even at old ages, or as yet unknown evolutionary effects due to the
presence of a companion white dwarf (albeit at a large distance from
the M dwarf).

We have successfully extended the activity$-$age relations to ages
several Gyr older than the cluster relations, but it is clear that
more work needs to be done to increase the number of old M dwarf stars
for which accurate ages can be determined.  For a more complete
picture of old M stars, work must also be done to extend this relation
to later spectral types, as our study is not well sampled at types
later than M4.5. Recent studies \citep{West} show that magnetic
activity is strongly correlated with spectral type in large samples of
disk stars, and also depends on distance from the Galactic plane,
another age-dependent factor.

In the context of previous work, our analysis suggests that nearly all
of our high velocity common proper motion binary WDs are part of a
rotating component and have kinematics which resemble the high
velocity tail of the Galaxy's disk population as found in SOH.  Also,
the M dwarf stars in our sample exhibit metallicities indicative of
the thick disk component of the Galaxy, just as \citet{RSH}
demonstrated with their large sample of M dwarf stars.  Our study of
WD+dM systems demonstrates that the active M dwarfs exhibit mostly
thin disk kinematics, as expected of a younger population. The
metallicities afforded by M dwarf companions provide essential
confirmation of population membership.

\acknowledgments

This work was supported by the NASA Graduate Researchers Program Grant
NGT 200415 (NMS); NFS AST02-05875 (NMS, SLH); A Grant-in-Aid of
Research from the National Academy of Sciences administered by Sigma
Xi, The Scientific Research Society (NMS); NASA Grant Y701296 (TDO);
and NSF Grant AST 0206115 (TDO).

\clearpage

%%%%%%%%%%%%%%%%%%%%%%%%%%%%  TABLES  %%%%%%%%%%%%%%%%%%%%%%%%%%%%%%%%%%%%%%%
%% Table 1 %%%%%%%%%%%%%%%%%%%%%%%%%%%%%%%%%%%%%%%%%%%%%%%%%%%%%%%%%%%%%%%%%%

\LongTables
\begin{landscape}

\begin{deluxetable}{lrrclclllrcrrr}
\tabletypesize{\small}
\setlength{\tabcolsep}{0.04in}
%\rotate
\tablecolumns{14}
\tablewidth{0pt}
\tablecaption{Complete M Dwarf$-$White Dwarf Wide Binary Target List. \label{all}}
\tablehead{
\colhead{Identifier} & \colhead{R.A.} & \colhead{Decl.} & \colhead{$V$1} & \colhead{Sp.1\tablenotemark{a}} &
\colhead{$V$2} & \colhead{Sp.2\tablenotemark{a}} & \colhead{Site} & \colhead{U.T.} & \colhead{Exp.} &
\colhead{$\mu$} & \colhead{$\Theta_{\mu}$} & \colhead{Pos.} & \colhead{Sep.}\\
\colhead{(Sp.1/Sp.2)} & \colhead{(B1950)} & \colhead{(B1950)} & \colhead{} & \colhead{} &
\colhead{} & \colhead{} & \colhead{} & \colhead{(mm/yy)} & \colhead{(sec)} &
\colhead{($\arcsec$/yr)} & \colhead{(deg)} & \colhead{(deg)} & \colhead{($\arcsec$)}\\ 
\colhead{(1)} & \colhead{(2)} & \colhead{(3)} & \colhead{(4)} & \colhead{(5)} & \colhead{(6)} &
\colhead{(7)} & \colhead{(8)} & \colhead{(9)} & \colhead{(10)} & \colhead{(11)} & 
\colhead{(12)} & \colhead{(13)} & \colhead{(14)}} 
\startdata
LP349$-$14/13   & 00:22:06 & 27:24:00  & 13.40    & dM   & 16.86    & DB  & APO  & 06/01 & 300 & 0.12 & 90 & 305 & 113 \\
G158$-$77/78    & 00:23:28 & -10:53:30 & 13.47    & sdM1 & 16.23    & DA  & APO  & 09/00,06/01 & 300 & 0.27 & 160 & 127 & 90\\
G171$-$B10A/B   & 00:23:53 & 38:52:24  & 15.97    & DA   & 15.26    & dM5 & APO  & 09/00 & 900 & 0.20 & 132 & 202 & 25 \\
L170$-$14A/B    & 00:27:30 & -54:58:00 & 14.84    & dM3.5 & 16.23   & DA  & CTIO & 01/02 & 900 & 0.36 & 240 & 327 & 3 \\
G218$-$8/7      & 00:38:30 & 55:33:42  & 14.1\phn & DQ5  & 14.02    & dM  & APO  & 09/00 & 1000 & 0.32 & 103 & 246 & 11\\
LP825$-$712/713 & 00:40:36 & -22:02:00 & 15.7\phn & sdM1 & 18.0\phn & DA3 & APO  & 09/01 & 900 & 0.50 & 83 & 50 & 23\\
LP242$-$11/12   & 00:52:30 & 38:32:00  & 13.80    & dM0  & 18.29    & DC  & APO  & 09/00,09/01 & 300 & 0.50 & 242 & 56 & 22\\
LP1$-$169/170   & 01:01:00 & 86:40:00  & 15.2\phn & dM1  & 16.7\phn & DQ  & APO  & 12/00 & 450 & 0.13 & 73 & 326 & 7 \\
\enddata
\tablenotetext{a}{Spectral types for the WD and dM stars were determined from low-resolution ($\sim$7-15 \AA) spectra \citep{OHL,OSHL,O93}.
Improved dM spectral types are given in Table \ref{ew}.  We have made no attempt to improve upon the WD spectral types in this study.}
\tablecomments{Single-precision $V$ magnitude values are $m_{\rm pg}$ magnitudes from Luyten's original survey, all others are 
$V$ magnitudes from Smith (1997).}
\end{deluxetable}

\clearpage 
\end{landscape}

%% Table 2 %%%%%%%%%%%%%%%%%%%%%%%%%%%%%%%%%%%%%%%%%%%%%%%%%%%%%%%%%%%%%%%%%%%%%%

\begin{deluxetable}{ccclll}
\tabletypesize{\small}
\setlength{\tabcolsep}{0.04in}
\tablecolumns{6}
\tablewidth{0pt}
\tablecaption{Journal of the Observations. \label{lo}}
\tablehead{
\colhead{}     & \colhead{U.T.} & \colhead{}      & \colhead{Site} & \colhead{Seeing}   & \colhead{Notes} \\
\colhead{(yr)} & \colhead{(mo)} & \colhead{(day)} & \colhead{}     & \colhead{(arcsec)} & \colhead{}\\
\colhead{(1)}  & \colhead{(2)}  & \colhead{(3)}   & \colhead{(4)}  & \colhead{(5)}      & \colhead{(6)}}
\startdata
2000 & Sep & 04 & APO  & $\sim$1.0 & gibbous moon, clear \\
2000 & Sep & 05 & APO  & 1.0       & gibbous moon, clear \\
2000 & Sep & 06 & APO  & 1.0       & full moon, clear \\   
2000 & Sep & 07 & APO  & 0.8       & partly cloudy \\      
2000 & Oct & 03 & APO  & 0.8       & partly cloudy \\   
2000 & Dec & 06 & APO  & $\sim$1.0 & clear \\
2001 & Jan & 24 & APO  & $\sim$3.2 & high humidity, variable seeing \\
2001 & Jan & 25 & APO  & 2.7       & high cirrus and humidity \\
\enddata
\end{deluxetable}

\clearpage 

%% Table 3 %%%%%%%%%%%%%%%%%%%%%%%%%%%%%%%%%%%%%%%%%%%%%%%%%%%%%%%%%%%%%%%%%%%

\LongTables             
\begin{landscape}

\begin{deluxetable}{lcrccccccccccccc}
\tabletypesize{\small}
\setlength{\tabcolsep}{0.04in}
%\rotate
\tablecolumns{16}
\tablewidth{0pt}
\tablecaption{Equivalent Widths and BandHead Flux Ratios. \label{ew}}
\tablehead{
\colhead{Identifier} & \colhead{R.A.} & \colhead{Decl.} & \colhead{EW H$\alpha$} & \colhead{$\sigma_{H\alpha}$} & 
\colhead{$\sigma_{H\alpha}$} & \colhead{CaH1} & \colhead{CaH2} & \colhead{CaH3} & \colhead{TiO1} & 
\colhead{TiO2} & \colhead{TiO3} & \colhead{TiO4} & \colhead{TiO5} & \colhead{Sp.\tablenotemark{a}} & \colhead{Ref.} \\
\colhead{(dM)} & \colhead{(B1950)} & \colhead{(B1950)} & \colhead{(6563\AA)} & \colhead{(+ \AA)} & \colhead{($-$ \AA)} &
\colhead{(6385\AA)} & \colhead{(6830\AA)} & \colhead{(6975\AA)} & \colhead{(6720\AA)} & \colhead{(7059\AA)} &
\colhead{(7094\AA)} & \colhead{(7132\AA)} & \colhead{(7050\AA)} & \colhead{(dM)} & \colhead{} \\
\colhead{(1)} & \colhead{(2)} & \colhead{(3)} & \colhead{(4)} & \colhead{(5)} & \colhead{(6)} & \colhead{(7)} &
\colhead{(8)} & \colhead{(9)} & \colhead{(10)} & \colhead{(11)} & \colhead{(12)} & \colhead{(13)} & \colhead{(14)} & 
\colhead{(15)} & \colhead{(16)}} 
\startdata
LP349$-$14  & 00:22:06 &  27:24:00 & -0.29 & 0.02 & 0.02 & 0.85 & 0.59 & 0.81 & 0.88 & 0.81 & 0.86 & 0.81 & 0.66 & M1.0 & 3 \\
G158$-$77   & 00:23:28 & -10:53:30 & -0.39 & 0.02 & 0.00 & 0.91 & 0.83 & 0.92 & 0.93 & 0.91 & 0.93 & 0.87 & 0.79 & K7.5 & 3 \\
G171$-$B10B & 00:23:53 &  38:52:24 & -0.10 & 0.05 & 0.00 & 0.84 & 0.43 & 0.73 & 0.80 & 0.58 & 0.71 & 0.64 & 0.39 & M4.0 & 3 \\
L170$-$14A  & 00:27:30 & -54:58:00 &  \phn 3.33 & 0.02 & 0.26 & 0.86 & 0.44 & 0.71 & 0.81 & 0.48 & 0.67 & 0.61 & 0.34 & M4.5 & 1 \\
G218$-$7    & 00:38:30 &  55:33:42 & -0.10 & 0.01 & 0.03 & 0.82 & 0.44 & 0.72 & 0.76 & 0.60 & 0.67 & 0.60 & 0.39 & M4.0 & 3 \\
Wolf 53     & 01:02:51 &  61:04:18 & -0.03 & 0.00 & 0.02 & 0.76 & 0.49 & 0.73 & 0.78 & 0.69 & 0.75 & 0.69 & 0.47 & M3.0 & 3 \\
LP587$-$54  & 01:20:42 & -02:24:48 & -0.31 & 0.01 & 0.01 & 0.87 & 0.58 & 0.80 & 0.84 & 0.76 & 0.80 & 0.75 & 0.57 & M2.0 & 3 \\
G34$-$48    & 01:42:31 &  23:02:48 & -0.19 & 0.00 & 0.01 & 0.86 & 0.65 & 0.82 & 0.88 & 0.82 & 0.85 & 0.81 & 0.67 & M0.5 & 3 \\
\enddata
\tablenotetext{a}{Improved dM spectral types based on TiO5 bandhead ratios.}
\tablenotetext{b}{Stellar halo white dwarf candidate.}
\tablenotetext{c}{M dwarf has close (eclipsing) companion.} 
\tablerefs{(1) dMe Stars; (2) dM(e) Stars; (3) dM Stars.}
\end{deluxetable}

\clearpage
\end{landscape}

%% Table 4 %%%%%%%%%%%%%%%%%%%%%%%%%%%%%%%%%%%%%%%%%%%%%%%%%%%%%%%%%%%%%%%%%%%%%

\begin{deluxetable}{lcrccc}
\tabletypesize{\small}
\setlength{\tabcolsep}{0.04in}
\tablecolumns{6}
\tablewidth{0pt}
\tablecaption{Parameters for the dMe and dM(e) Stars. \label{vimv}}
\tablehead{
\colhead{Identifier} & \colhead{R.A.} & \colhead{Decl.} &
\colhead{$V-I$} & \colhead{$M_{V}$} & \colhead{Mass} \\
\colhead{(dM)} & \colhead{(B1950)} & \colhead{(B1950)} & \colhead{} & 
\colhead{} & \colhead{(M$_{\odot}$)} \\
\colhead{(1)} & \colhead{(2)} & \colhead{(3)} & \colhead{(4)} & 
\colhead{(5)} & \colhead{(6)}}
\startdata
\cutinhead{dMe Stars}                   
L170$-$14A  & 00:27:30 & -54:58:00 & 3.30 & 13.51 & 0.18 \\
LP31$-$140  & 03:24:30 &  73:15:00 & 3.09 & 13.01 & 0.21 \\
HZ 9B       & 04:29:24 &  17:38:00 & 2.52 & 11.34 & 0.34 \\
LP891$-$13  & 04:43:18 & -27:32:00 & 2.86 & 13.00 & 0.21 \\
LP205$-$28  & 06:41:50 &  43:48:36 & 2.59 & 11.55 & 0.32 \\
LP313$-$13  & 09:04:30 &  27:49:00 & 3.07 & 12.94 & 0.21 \\
LP487$-$26  & 09:16:00 &  11:11:00 & 3.23 & 13.35 & 0.19 \\
G117$-$B15B & 09:21:13 &  35:29:48 & 2.77 & 12.11 & 0.27 \\
\enddata
\end{deluxetable}

\clearpage

%% Table 5 %%%%%%%%%%%%%%%%%%%%%%%%%%%%%%%%%%%%%%%%%%%%%%%%%%%%%%%%%%%%%%%%%%%%%%

\LongTables             
\begin{landscape}

\begin{deluxetable}{lcrcrrrrrrrrrrrr}
\tabletypesize{\small}
\setlength{\tabcolsep}{0.04in}
\tablecolumns{16}
%\rotate
\tablewidth{0pt}
\tablecaption{Complete Space Motions for 161 M Dwarf Stars. \label{uvw}}
\tablehead{
\colhead{Identifier} & \colhead{R.A.} & \colhead{Decl.} & \colhead{$\mu$} & \colhead{$\Theta$} & \colhead{$v_{r}$} & 
\colhead{$\sigma_{v_{r}}$} & \colhead{$U$} & \colhead{$\sigma_{U}$} & \colhead{$V$} & \colhead{$\sigma_{V}$} & \colhead{$W$} & 
\colhead{$\sigma_{W}$} & \colhead{Sp.V.} & \colhead{$\sigma_{\rm Sp.V.}$} & \colhead{Ref.} \\
\colhead{(dM)} & \colhead{(B1950)} & \colhead{(B1950)} & \colhead{($\arcsec$/yr)} & \colhead{(deg)} & \colhead{(km/s)} &
\colhead{(km/s)} & \colhead{(km/s)} & \colhead{(km/s)} & \colhead{(km/s)} & \colhead{(km/s)} & \colhead{(km/s)} &
\colhead{(km/s)} & \colhead{(km/s)} & \colhead{(km/s)} & \colhead{} \\
\colhead{(1)} &  \colhead{(2)} & \colhead{(3)} & \colhead{(4)} & \colhead{(5)} & \colhead{(6)} & \colhead{(7)} & 
\colhead{(8)} & \colhead{(9)} & \colhead{(10)} & \colhead{(11)} & \colhead{(12)} & \colhead{(13)} & \colhead{(14)} & 
\colhead{(15)} & \colhead{(16)}}
\startdata
LP349$-$14  & 00.3683 &  27.4000 & 0.12 &  90 &  48.6 &  3.5 &  37.2 & 0.9 & 7.4 & 1.9 & -38.2 & 2.8 & 53.8 & 0.1 & 3 \\
G158$-$77   & 00.3911 & -10.8917 & 0.27 & 160 &  52.2 &  6.4 & -27.6 & 0.9 & -110.5 & 3.2 & -93.5 & 5.5 & 147.3 & 0.0 & 3 \\
G171$-$B10B & 00.3981 &  38.8733 & 0.20 & 132 &  19.4 &  2.9 &  20.1 & 0.9 & -22.5 & 1.8 & -46.0 & 2.1 & 55.0 & 0.0 & 3 \\
L170$-$14A  & 00.4583 & -54.8917 & 0.36 & 240 &  16.2 & 14.8 & -43.8 & 6.0 & -16.4 & 9.9 & -11.9 & 9.2 & 48.2 & 0.2 & 1 \\
G218$-$8    & 00.6417 &  55.5589 & 0.32 & 103 &  20.5 &  8.5 &  21.2 & 3.5 & -7.5 & 5.7 & -15.9 & 5.2 & 27.5 & 0.2 & 3 \\
LP825$-$712 & 00.6767 & -22.0333 & 0.50 &  83 &  11.9 & 30.0 &  53.1 & 7.2 & -39.8 & 16.0 & -20.3 & 24.3 & 69.4 & 0.3 & 4 \\
LP242$-$11  & 00.8750 &  32.5333 & 0.50 & 242 &  22.8 & 33.6 & -159.5 & 10.6 & 58.1 & 19.3 & -96.8 & 25.4 & 195.4 & 0.1 & 2,4 \\
LP1$-$169   & 01.0167 &  86.6667 & 0.13 &  73 &  23.7 &  5.1 &  11.6 & 2.5 & -1.7 & 3.8 & 6.5 & 2.4 & 13.4 & 0.3 & 2,4 \\
\enddata
\tablecomments{Units of right ascension are hours, and units of declination are degrees.}
\tablerefs{(1) dMe Stars; (2) dM(e) Stars; (3) dM Stars; (4) $V$ and $M_{V}$ are from the WD companion.}
\tablenotetext{a}{Stellar halo white dwarf candidate.}
\tablenotetext{b}{M dwarf has close (eclipsing) companion.} 
\end{deluxetable}

\end{landscape}

\end{document}